\documentclass[12pt]{iopart}
\usepackage{iopams}
\usepackage{setstack}

\newcommand{\half}{\textstyle{\frac{1}{2}}}
\newcommand{\quarter}{\textstyle{\frac{1}{4}}}
\newcommand{\threequarter}{\textstyle{\frac{3}{4}}}
\newcommand{\third}{\textstyle{\frac{1}{3}}}

\newcommand{\cP}{{\cal P}}
\newcommand{\cT}{{\cal T}}

\begin{document}

\title[Quantum effects in classical systems having complex energy]
{Quantum effects in classical systems having complex energy}

\author[Bender, Brody, and Hook]{Carl~M~Bender${}^\ast$, Dorje~C~Brody${}
^\dag$, and Daniel~W~Hook${}^\ddag$}

\address{${}^\ast$Department of Physics, Washington University, St. Louis, MO
63130, USA \\{\footnotesize{\tt email: cmb@wustl.edu}}}

\address{${}^\dag$Department of Mathematics, Imperial College London, London
SW7 2AZ, UK\\ {\footnotesize{\tt email: dorje@imperial.ac.uk}}}

\address{${}^\ddag$Blackett Laboratory, Imperial College London, London SW7 2AZ,
UK\\ {\footnotesize{\tt email: d.hook@imperial.ac.uk}}}

\date{today}

\begin{abstract}
On the basis of extensive numerical studies it is argued that there are strong 
analogies between the probabilistic behavior of quantum systems defined by
Hermitian Hamiltonians and the deterministic behavior of classical mechanical
systems extended into the complex domain. Three models are examined: the quartic
double-well potential $V(x)=x^4-5x^2$, the cubic potential $V(x)=\half x^2-gx^3
$, and the periodic potential $V(x)=-\cos x$. For the quartic potential a wave
packet that is initially localized in one side of the double-well can tunnel to
the other side. Complex solutions to the classical equations of motion exhibit a
remarkably analogous behavior. Furthermore, classical solutions come in two
varieties, which resemble the even-parity and odd-parity quantum-mechanical
bound states. For the cubic potential, a quantum wave packet that is initially
in the quadratic portion of the potential near the origin will tunnel through
the barrier and give rise to a probability current that flows out to infinity.
The complex solutions to the corresponding classical equations of motion exhibit
strongly analogous behavior. For the periodic potential a quantum particle whose
energy lies between $-1$ and $1$ can tunnel repeatedly between adjacent
classically allowed regions and thus execute a localized random walk as it hops
from region to region. Furthermore, if the energy of the quantum particle lies
in a conduction band, then the particle delocalizes and drifts freely through
the periodic potential. A classical particle having complex energy executes a
qualitatively analogous local random walk, and there exists a narrow energy band
for which the classical particle becomes delocalized and moves freely through
the potential.

\end{abstract}

\submitto{\JPA}

\section{Introduction}
\label{s1}
Quantum mechanics and classical mechanics provide profoundly different
descriptions of the physical world. In one-dimensional classical mechanics one
is given a Hamiltonian of the form $H(x,p)=\half p^2+V(x)$. The motion of a
particle modeled by this Hamiltonian is deterministic and is described by
Hamilton's equations
\begin{equation}
{\dot x}=\frac{\partial H}{\partial p}=p,\qquad{\dot p}=-\frac{\partial H}{
\partial x}=-V'(x),
\label{e1}
\end{equation}
or equivalently, by Newton's law $-V'(x)={\ddot x}$. The position $x(t)$ of a
particle at time $t$ is found by solving a {\it local} initial-value problem for
these differential equations. The energy $E$ of a particle, that is, the
numerical value of the Hamiltonian, is a constant of the motion and can take on
continuous values. Particle motion is restricted to the classically allowed
regions, which are defined by $E\geq V(x)$. Because a particle may not enter a
classically forbidden region, where $E<V(x)$, a classical particle may not
travel between disconnected classically allowed regions.

In quantum mechanics Heisenberg's operator equations of motion and the
time-dependent Schr\"odinger equation are posed as initial-value problems, just
as Hamilton's equations are treated as initial-value problems in classical
mechanics. However, the time-dependent Schr\"odinger equation is required to
satisfy {\it nonlocal} boundary conditions that guarantee that the total
probability of finding the particle is finite. For stationary states these
boundary conditions demand that the eigenfunctions of the Hamiltonian satisfy a
nonlocal boundary-value problem. As a consequence, for a rising potential that
confines a classical particle the energy spectrum is discrete.

While classical mechanics consists of nothing more than solving a differential
equation to find the exact trajectory of a particle, quantum mechanics is an
abstract theory in which the physical state of the system is represented by a
vector in a Hilbert space and predictions are probabilistic. The nonlocality
mentioned above implies that physical measurements are subtle and difficult to
perform. Quantum effects such as discretized energies, tunneling, and
interference are a consequence of the nonlocal nature of the theory and are not
intuitive. For example, one cannot speak of an actual path that a particle
follows when it tunnels from one classically allowed region to another.

Complex-variable theory is of great assistance in providing an understanding of
nonintuitive real-variable phenomena. It explains, for example, why the Taylor
series for a real function $f(x)$ may cease to converge at a real value of $x$
where $f(x)$ is smooth. (Series convergence is linked to the presence of
singularities that may lie in the complex plane and not on the real axis.)
Moreover, complex analysis shows the fundamental theorem of algebra,
which is a deep property of the roots of polynomials, to be nothing more than a
straightforward application of Liouville's theorem.

The objective of this conjectural paper is to demystify some well-known quantum
effects by showing that their qualitative features can be reproduced very simply
by the deterministic equations of classical mechanics (Newton's law) when these
equations are extended to and solved in the complex plane. Specifically, we take
the uncertainty principle $\Delta E\,\Delta t\gtrsim\half\hbar$ to mean that
there is intrinsic uncertainty in the energy of a particle, and in this paper we
consider the possibility that this uncertainty may have an imaginary as well as
a real part. We find that a deterministic classical particle whose energy has a
small imaginary component can exhibit phenomena that are associated exclusively
with quantum mechanics. We do not necessarily claim that quantum mechanics is a
deterministic hidden-complex-variable theory. Indeed, there are important
quantum phenomena, such as interference effects, that we cannot as yet reproduce
by using complex classical mechanics. However, the results that we obtain by
using complex classical mechanics to simulate quantum mechanics bear a striking
qualitative and quantitative resemblance to many well-known quantum effects.

This paper is organized as follows: In Sec.~\ref{s2} we illustrate the power of
complex analysis by using it to explain the quantization of energy. We show that
in the complex domain the energy levels cease to be discrete and energy
quantization can be explained topologically as the counting of sheets of a
Riemann surface. In Sec.~\ref{s3} we describe the general features of classical
particle trajectories when the energy of the classical particle is allowed to be
complex. Specifically, classical trajectories that are closed and periodic when
the energy is real cease to be closed when the energy becomes complex. In
Secs.~\ref{s4}, \ref{s5}, and \ref{s6} we examine the complex particle
trajectories for three potentials whose quantum properties are well studied:
the quartic double-well potential $x^4-5x^2$, the cubic potential $x^2-gx^3$,
and the periodic potential $-\cos x$. In each of these cases, we find that the
corresponding complex classical system is able to mimic the quantum phenomena of
tunneling, bound states of distinct parity, conduction bands, and energy gaps.
Section \ref{s7} contains some concluding remarks.

\section{Quantization from the complex-variable perspective}
\label{s2}

The notion of quantized energy levels is a central feature of quantum mechanics
and is a dramatic departure from the continuous energy associated with classical
mechanics. One can gain a different perspective on quantization if one extends
quantum theory into the complex domain. To illustrate this, we consider a simple
two-dimensional quantum system whose Hamiltonian is
\begin{equation}
H=H_0+\epsilon H_I,
\label{e2}
\end{equation}
where the coupling constant $\epsilon$ is real. The diagonal matrix
\begin{equation}
H_0=\left(\begin{array}{cc}a&0\\0&b\end{array}\right),
\label{e3}
\end{equation}
whose two energy levels are $a$ and $b$, describes the unperturbed system. The
interaction is represented by the Hermitian off-diagonal matrix
\begin{equation}
H_I=\left(\begin{array}{cc}0&c\\c&0\end{array}\right).
\label{e4}
\end{equation}
The energy levels of $H$ are evidently real and discrete:
\begin{equation}
E_\pm=\half\left[a+b\pm\sqrt{(a-b)^2+4\epsilon^2c^2}\right].
\label{e5}
\end{equation}

\begin{figure*}[t!]
\vspace{2.5in}
\includegraphics{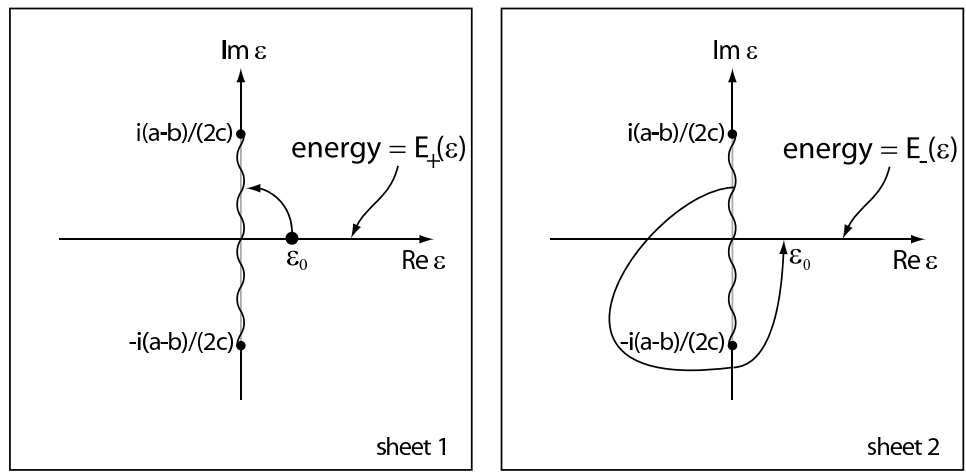}
\caption{Two-sheeted Riemann surface for the energy function $E(\epsilon)$ in
(\ref{e6}). On the Riemann surface this function is smooth and continuous, and
the quantization of energy levels corresponds to counting the sheets on the
surface. The energy appears to be discrete and quantized only if we limit
the Riemann surface to the real-$\epsilon$ axis. A path from $\epsilon_0$ on
the real axis of sheet 1 to the corresponding point on the real axis of sheet 2
is shown. Along this path the energy eigenvalue $E_+$ continuously deforms to
the other energy eigenvalue $E_-$.}
\label{f1}
\end{figure*}

An elementary way to understand the discreteness of these energy levels is to
extend the coupling constant $\epsilon$ into the complex domain: Define the {\it
energy function} $E(\epsilon)$ by
\begin{equation}
E(\epsilon)\equiv\half\left[a+b+\sqrt{(a-b)^2+4\epsilon^2c^2}\right].
\label{e6}
\end{equation}
As a function of complex $\epsilon$, $E(\epsilon)$ is double-valued, so it must
be defined on a two-sheeted Riemann surface. These sheets are joined at a branch
cut that connects the square-root branch points located at $\epsilon=\pm i(a-b)/
(2c)$ (see Fig.~\ref{f1}). On the real-$\epsilon$ axis of the first sheet $E(
\epsilon)=E_+$, and on the real-$\epsilon$ axis of the second sheet $E(\epsilon)
=E_-$. On the Riemann surface the energy function $E(\epsilon)$ is smooth and
{\it continuous} and is not a quantized function of complex $\epsilon$. Indeed,
along a continuous path that runs from a point $\epsilon_0$ on the real axis on
the first sheet, crosses the branch cut, and goes to the corresponding point
$\epsilon_0$ on the real axis on the second sheet, the energy eigenvalue $E_+$
continuously deforms to $E_-$. (Such a path is shown on Fig.~\ref{f1}.) Thus,
we see that the quantization in (\ref{e5}) is a consequence of the topological
discreteness of the sheets that make up the Riemann surface. The energy function
$E(\epsilon)$ appears quantized only if we restrict its domain to the real axes
on the sheets of the Riemann surface.

To summarize, we have extended the Hamiltonian in (\ref{e2}) into the complex
domain by complexifying the coupling constant $\epsilon$ and have obtained a
clearer and deeper understanding of the nature of quantization. The topological
picture of quantization described here is quite general, and it applies to more
complicated systems, such as $H=p^2+V(x)+\epsilon W(x)$, which have an infinite
number of energy levels \cite{R1}.

In general, the advantage of analyzing a system in the complex plane is that
special features (like the discreteness of eigenvalues), which only occur on the
real axis or which only emerge when we limit our attention to the real domain,
can be seen to be part of a simpler and more general framework. In the rest of
this paper we will examine complexified classical mechanics. We will see that
while classical trajectories tend to be closed and periodic when the energy is
strictly real, this special feature disappears and trajectories cease to be
closed when the energy is allowed to be complex. Open trajectories are generic
and, unlike closed trajectories, their behavior is rich and elaborate and bears
a strong resemblance to some of the features that are thought to be restricted
to the domain of quantum mechanics.

\section{Classical mechanics in the complex domain}
\label{s3}

Given a classical Hamiltonian $H(x,p)$, the path $x(t)$ of a particle is fully
determined by Hamilton's equations of motion (\ref{e1}) together with the
initial conditions $x(0)$ and $p(0)$. The energy $E$ is fixed by these initial
conditions and is left invariant under the action of Hamilton's equations of
motion. In elementary texts on classical mechanics the initial conditions are
taken to be real so that the energy $E$ is real and, in addition, particle
trajectories are restricted to the real-$x$ axis. However, in recent papers on
$\cP\cT$-symmetric classical mechanics, it has been shown that it is interesting
to study the complex as well as the real trajectories for systems having real
energy \cite{R2,R3,R4,R5,R6,R7,R8,R9,R10,R11}.

\begin{figure*}[t!]
\vspace{2.6in}
\includegraphics{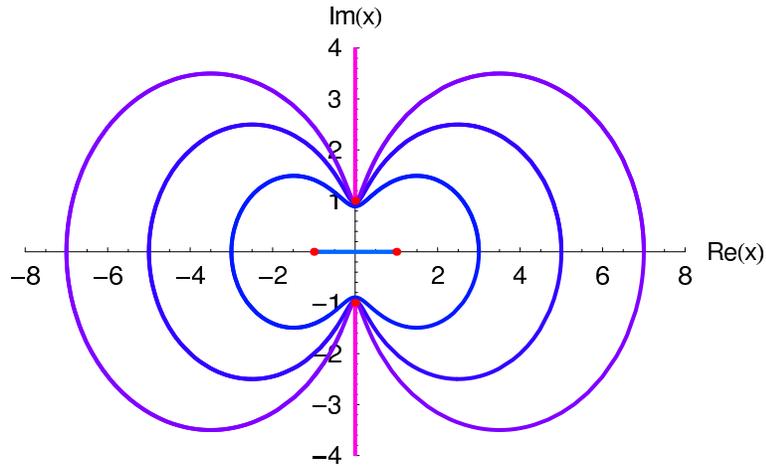}
\caption{Classical trajectories in the complex-$x$ plane representing the
possible motions of a particle of energy $E=1$. This motion is governed by the
anharmonic-oscillator Hamiltonian $H=\half p^2+x^4$. There is one real
trajectory that oscillates between the turning points at $x=\pm1$ and an
infinite family of nested complex trajectories that enclose the real turning
points but lie inside the imaginary turning points at $\pm i$. (The turning
points are indicated by dots.) Two other trajectories begin at the imaginary
turning points and drift off to infinity along the imaginary-$x$ axis. Apart
from the trajectories beginning at $\pm i$, all trajectories are closed and
periodic. All orbits in this figure have the same period $\sqrt{\pi/2}\,\Gamma
\left(\quarter\right)/\Gamma\left(\threequarter\right)=3.70815\ldots$.}
\label{f2}
\end{figure*}

We illustrate the real-energy trajectories of a classical-mechanical system by
using the anharmonic oscillator, whose Hamiltonian is $H=\half p^2+x^4$. The
classical trajectories for the energy $E=1$ are shown in Fig.~\ref{f2}. There
are four turning points located at $x=\pm1,~\pm i$ and indicated by dots. The
so-called ``classically allowed'' region is the portion of the real axis between
$x=-1$ and $x=1$, and a classical particle that is initially on this line
segment will move parallel to the real axis and oscillate between the real
turning points. The so-called ``classically forbidden'' regions are the portions
of the real-$x$ axis for which $|x|>1$, and a particle whose initial position is
in one of these regions will have an initial motion that is perpendicular to the
real axis. The particle will then enter the complex plane, make a sharp turn
about the imaginary turning points, and return to its initial position. By
virtue of Cauchy's theorem, all closed orbits in this figure have the same
period $\sqrt{\pi/2}\,\Gamma\left(\quarter\right)/\Gamma\left(\threequarter
\right)=3.70815\ldots$. There are two open orbits that run along the imaginary
axis from $i$ to $+i\infty$ and from $-i$ to $-i\infty$ in half this time. Note
that two different classical trajectories can never cross.

The crucial feature illustrated in Fig.~\ref{f2} is that all of the classical
trajectories (except the two that run off to infinity along the imaginary axis)
are closed. This means that we can view the system as a sort of complex atom.
Because the classical orbits are closed, we can quantize the system and
calculate the allowed real energies $E_n$ by using the Bohr-Sommerfeld
quantization formula $\oint dx\,p=(n+\half)\pi$ along {\it any} of these closed
orbits to obtain the real discrete energy levels of the quantum anharmonic
oscillator.

\begin{figure*}[t!]
\vspace{2.1in}
\includegraphics{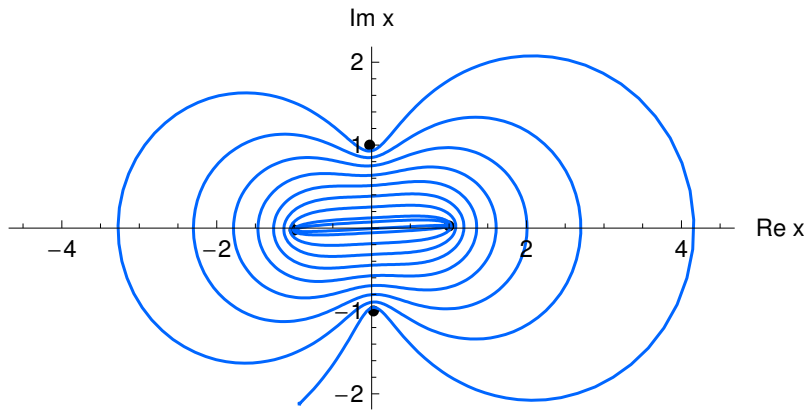}
\caption{A single classical trajectory in the complex-$x$ plane for a particle
governed by the anharmonic-oscillator Hamiltonian $H=\half p^2+x^4$. This
trajectory begins at $x=1$ and represents the complex path of a particle whose
energy $E=1+0.1i$ is complex. The trajectory is not periodic because it is not
closed. The four turning points are indicated by dots. The trajectory does not
cross itself.}
\label{f3}
\end{figure*}

Recall that the Heisenberg uncertainty principle states that there is an
intrinsic uncertainty in the energy $E$. Let us see what happens if this
uncertainty in energy implies that it can take complex values. To begin with,
since the turning points are determined by the value of the energy, they are
slightly displaced from their positions in Fig.~\ref{f2}. However, the main
effect is that while the classical trajectories still do not cross, they no
longer need be closed and periodic. In Fig.~\ref{f3} a single trajectory for a
particle whose energy is $E=1+0.1i$ is shown. The initial position is $x(0)=1$
and the particle is allowed to travel for a time $t_{\rm max}=35$.

\section{Double-well potential}
\label{s4}

In this section we examine the double-well potential $x^4-5x^2$. A
negative-energy quantum particle in such a potential tunnels back and forth
between the classically allowed regions to the left and to the right of $x=0$.

\begin{figure*}[t!]
\vspace{2.6in}
\includegraphics{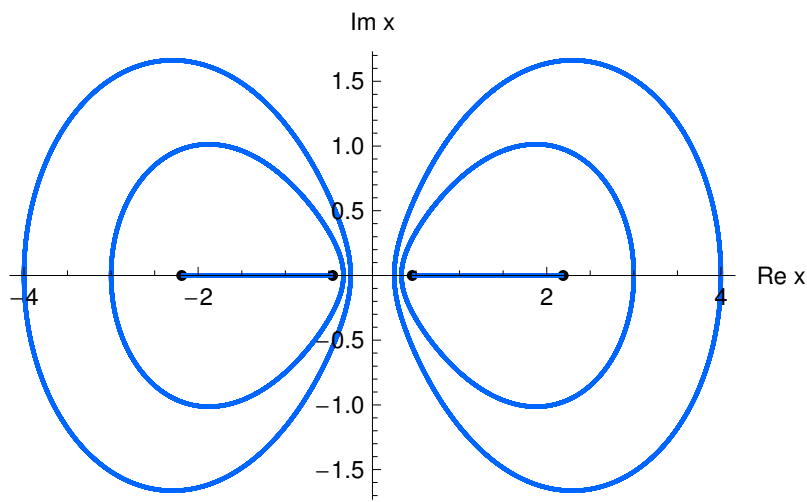}
\caption{Six classical trajectories in the complex-$x$ plane representing the
motion of a particle of energy $E=-1$ in the potential $x^4-5x^2$. The turning
points are located at $x=\pm2.19$ and $x=\pm0.46$ and are indicated by dots.
Because the energy is real, the trajectories are all closed. The classical
particle stays in either the right-half or the left-half plane and cannot cross
the imaginary axis. Thus, when the energy is real, there is no effect analogous
to tunneling.}
\label{f4}
\end{figure*}

Let us first see what happens if we put a classical particle whose energy is
real in such a potential well. We give an energy of $E=-1$ to this particle and
plot some of the possible complex trajectories in Fig.~\ref{f4}. The turning
points associated with this choice of energy lie on the real-$x$ axis at $x=\pm
2.19$ and $x=\pm0.46$ and are indicated by dots. Observe that the classical
trajectories are always confined to either the right-half or the left-half plane
and do not cross the imaginary axis. Therefore, when the energy is real there is
no effect analogous to quantum tunneling.

\begin{figure*}[t!]
\vspace{4.7in}
\includegraphics{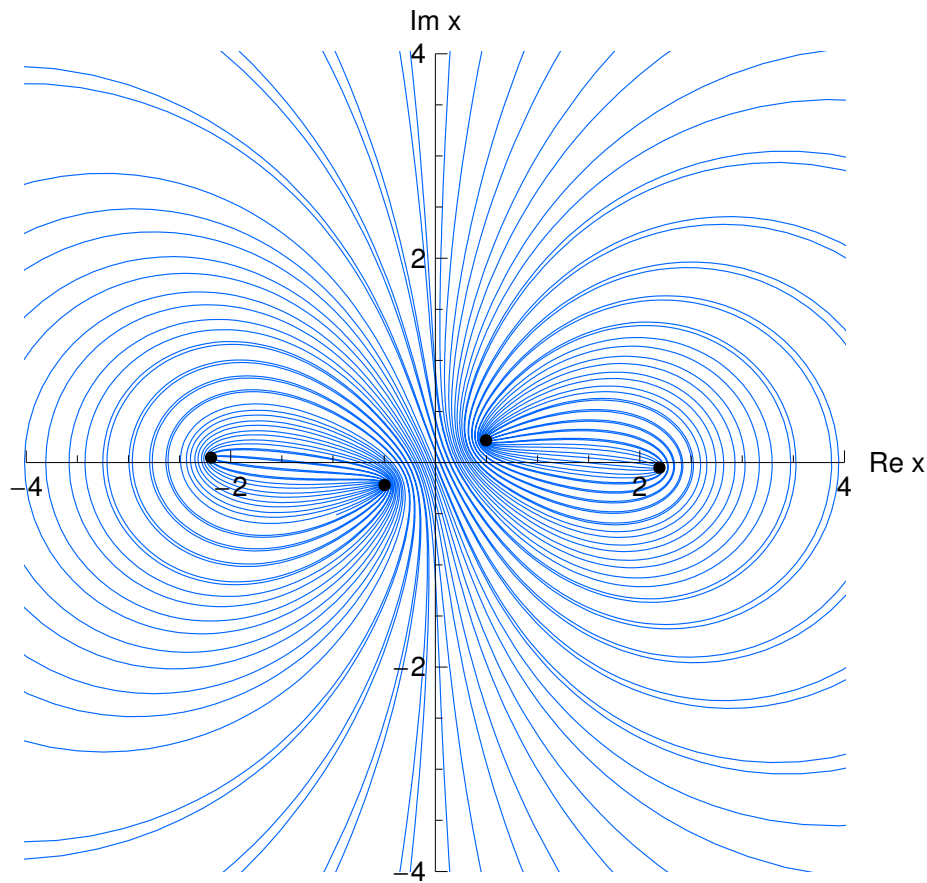}
\caption{Classical trajectory of a particle moving in the complex-$x$ plane
under the influence of a double-well $x^4-5x^2$ potential. The particle has
complex energy $E=-1-i$ and thus its trajectory does not close. The trajectory
spirals inward around one pair of turning points and then spirals outward. The
particle crosses the imaginary axis and spirals inward and then outward around
the other pair of turning points. It then crosses the imaginary axis and repeats
this behavior endlessly. At no point during this process does the trajectory
cross itself. This classical-particle motion is analogous to behavior of a
quantum particle that repeatedly tunnels back and forth between two
classically allowed regions. Here, however, the particle does not disappear into
the classically forbidden region during the tunneling process; rather, it moves
along a well-defined path in the complex-$x$ plane from one well to the other.
Table \ref{t1} shows that this trajectory and be thought of as having odd
parity.}
\label{f5}
\end{figure*}

Next, we allow the energy of the classical particle to be complex: $E=-1-i$. The
open classical trajectories that result from such a complex energy are
particularly interesting because their behavior is reminiscent of the phenomenon
of quantum tunneling. Figure \ref{f5} shows a single trajectory that begins at
$x=0$. The particle moves into the right-half complex plane, and as time passes,
the trajectory spirals inward around the right pair of turning points. The shape
of the spiral is similar to that in Fig.~\ref{f3}. After many turns, the
particle crosses the real axis between the two turning points and then begins to
spiral outward. Eventually, the particle crosses the imaginary axis and begins 
to spiral inward around the left pair of turning points. The process then
repeats: The particle eventually crosses the real axis, spirals outward, crosses
the imaginary axis, and begins to spiral inward around the right pair of turning
points.

This process of spiraling inward, spiraling outward, and crossing the imaginary
axis continues endlessly, but at no point does the trajectory ever cross itself.
During this process each pair of turning points acts like a strange attractor;
the pair of turning points draws the trajectory inward along a spiral path, but
then it drives the classical particle outward again along another nested spiral
path. The classical particle spends roughly half of its time spiraling under the
influence of the left pair of turning points and the other half of its time
spiraling under the influence of the right pair of turning points. The classical
``tunneling'' process is less abstract and hence easier to understand than its
quantum-mechanical analog. During quantum tunneling, the particle disappears
from one classical region and reappears almost immediately in another classical
region. We cannot ask which path the particle follows during this process.
However, for a classical particle, it is clear how the particle travels from one
classically allowed region to the other; it follows a well-defined path in the complex-$x$ plane.

\begin{table}[h]
\begin{center}
\begin{tabular}{|c|c|c|}\hline
Imaginary crossing point & Direction & Time of crossing \\\hline
$\;\;\;0.909\,592\,i$ & $\rightarrow$ & $45.728\,640$ \\
$\;\;\;0.781\,619\,i$ & $\leftarrow$ & $\;\;5.366\,490$ \\
$\;\;\;0.441\,760\,i$ & $\rightarrow$ & $34.347\,705$ \\
$\;\;\;0.407\,514\,i$ & $\leftarrow$ & $16.764\,145$ \\
$\;\;\;0.253\,436\,i$ & $\rightarrow$ & $22.909\,889$ \\
$\;\;\;0.231\,656\,i$ & $\leftarrow$ & $28.205\,183$ \\
$\;\;\;0.118\,499\,i$ & $\rightarrow$ & $11.457\,463$ \\
$\;\;\;0.100\,556\,i$ & $\leftarrow$ & $39.658\,755$ \\
$\;\;\;0\,i$ & $\rightarrow$ & $0$ \\
$-0.017\,057\,i$ & $\leftarrow$ & $51.116\,500$ \\
$-0.082\,877\,i$ & $\rightarrow$ & $90.775\,058$ \\
$-0.136\,772\,i$ & $\leftarrow$ & $62.573\,579$ \\
$-0.210\,728\,i$ & $\rightarrow$ & $79.320\,501$ \\
$-0.276\,231\,i$ & $\leftarrow$ & $74.024\,673$ \\
$-0.376\,304\,i$ & $\rightarrow$ & $67.876\,728$ \\
$-0.479\,881\,i$ & $\leftarrow$ & $85.458\,656$ \\
$-0.690\,666\,i$ & $\rightarrow$ & $56.467\,388$ \\
$-1.121\,155\,i$ & $\leftarrow$ & $96.812\,583$ \\
\hline
\end{tabular}
\end{center}
\caption{\label{t1} Imaginary-axis intercepts, directions, and times for the
classical trajectory shown in Fig.~\ref{f5}. Each time the trajectory crosses
the imaginary axis we register the intercept in the first column, the direction
of motion in the second column, and the time in the third column. This table
becomes its mirror image under spatial reflection, so we classify the
trajectory in Fig.~\ref{f5} as having odd parity.}
\end{table}

There is an even more surprising analogy between the quantum and classical
systems. A stationary state of a quantum particle in a double well like $x^4-5
x^2$ has a definite value of parity; that is, the eigenfunctions are either even
or odd functions of $x$. The classical trajectory shown in Fig.~\ref{f5} also
exhibits a sort of parity, which we can observe if we keep track of the
direction in which the particle is going when it crosses the imaginary axis. In
Table \ref{t1} we list in the first column the points at which the trajectory
crosses the imaginary axis and in the second column we indicate whether the
particle is crossing leftward or rightward. The third column indicates the time
of the crossing. Under parity reflection the directions of the velocities in the
second column reverse. However, the positions of the entries in the table must
also be reflected about the central entry because under parity we replace the
complex number $x$ by $a-x$ for some value of $a$. Thus, this table becomes its
mirror image under parity and we can classify the trajectory as having odd
parity.

\begin{figure*}[t!]
\vspace{4.7in}
\includegraphics{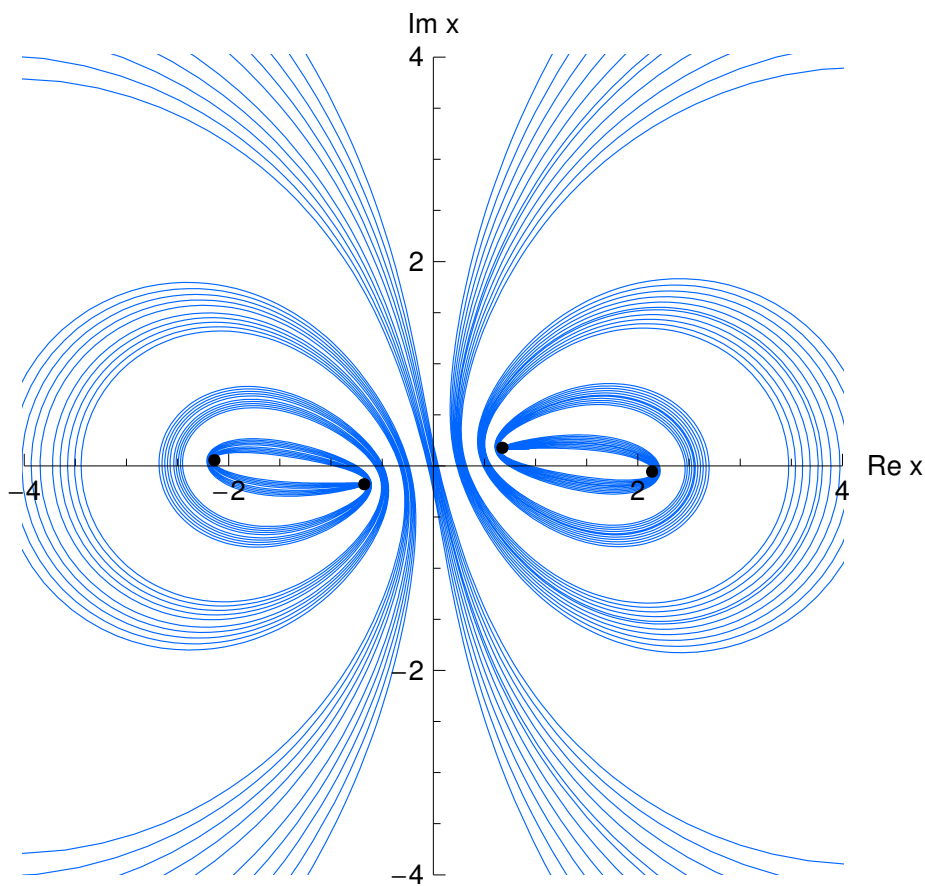}
\caption{Classical trajectory in the complex-$x$ plane for a particle in a
double-well $x^4-5x^2$ potential. The particle begins its motion at the origin
$x=0$ but has less real energy than the particle in Fig.~\ref{f5}: $E=-2-i$.
The trajectory is qualitatively different from that shown in Fig.~\ref{f5}
in that the path is confined to narrow ribbons and does not fill the complex-$x$
plane. Also, because the real part of the energy of the particle is less than
that for the particle in Fig.~\ref{f5}, the trajectory crosses the imaginary
axis (leaps over the barrier) less frequently. Table \ref{t2} of imaginary-axis
intercepts shows that we can interpret this trajectory as having even parity.}
\label{f6}
\end{figure*}

Let us now take a more negative value for the real part of the energy of the
classical particle while keeping the imaginary part of the energy the same: $E=-
2-i$. The particle trajectory shown in Fig.~\ref{f6} originates at $x=0$ and has
this energy. This trajectory is markedly different from that shown in
Fig.~\ref{f5} because the motion of the particle is confined to narrow bands or
ribbons. More importantly, when we construct a crossing table for this figure
(see Table \ref{t2}), we observe that the pattern of crossings is completely
different from that in the second column of Table \ref{t1}. In this case the
trajectory can be classified as having even parity.

\begin{table}[ht]
\begin{center}
\begin{tabular}{|c|c|c|}\hline
Imaginary crossing point & Direction & Time of crossing
\\\hline
$\;\;\;0.212\,966\,i$ & $\leftarrow$ & $85.604\,840$ \\
$\;\;\;0.114\,590\,i$ & $\leftarrow$ & $47.557\,393$ \\
$\;\;\;0.068\,159\,i$ & $\leftarrow$ & $28.534\,298$ \\
$\;\;\;0.407\,514\,i$ & $\leftarrow$ & $16.764\,145$ \\
$\;\;\;0.022\,623\,i$ & $\leftarrow$ & $9.511\,410$ \\
$\;\;\;0\,i$ & $\rightarrow$ & $0$ \\
$-0.045\,318\,i$ & $\rightarrow$ & $19.022\,837$ \\
$-0.091\,223\,i$ & $\rightarrow$ & $38.045\,810$ \\
$-0.187\,424\,i$ & $\rightarrow$ & $57.069\,065$ \\
$-0.210\,728\,i$ & $\rightarrow$ & $76.092\,765$ \\
$-0.239\,354\,i$ & $\rightarrow$ & $95.117\,102$ \\
\hline
\end{tabular}
\end{center}
\caption{\label{t2} Same as in Table \ref{t1}, but with entries corresponding to
the classical trajectory in Fig.~\ref{f6}. There are fewer crossings than in
Table \ref{t1} because the classical particle has less real energy. In
quantum-mechanical terms this means that the particle is less capable of leaping
over the barrier between the wells.}
\end{table}

The parities of the quantum eigenfunctions associated with the double-well
potential $x^4-5x^2$ alternate as the energy varies monotonically. Analogously,
we have found that the parities of the classical trajectories also alternate as
the real part of the energy changes monotonically.

\section{Cubic potential}
\label{s5}

In this section we examine the cubic Hamiltonian $H=\half p^2+\half x^2-gx^3$,
which is a model for a quantum particle in a long-lived metastable state. This
particle is initially confined to the classically allowed region in the
parabolic portion of the potential, but it eventually tunnels through the
barrier and then moves rapidly off to $x=+\infty$. This Hamiltonian serves as an
archetypal model for radioactive decay.

The energy levels of this Hamiltonian have been calculated for small $g$ by
using WKB theory \cite{R12}, and the approximate formula for the ground-state
energy is
\begin{equation}
E(g)\sim\half-\textstyle{\frac{11}{8}}g^2-i\textstyle{\frac{1}{g\sqrt{\pi}}}
e^{-2/(15g^2)}\qquad(g\to0^+).
\label{e7}
\end{equation}
The reciprocal of the imaginary part of the energy is an approximate measure of
the lifetime $\tau$ of the metastable state:
\begin{equation}
\tau\approx g\sqrt{\pi}e^{2/(15g^2)}.
\label{e8}
\end{equation}

\begin{figure*}[t!]
\vspace{2.8in}
\includegraphics{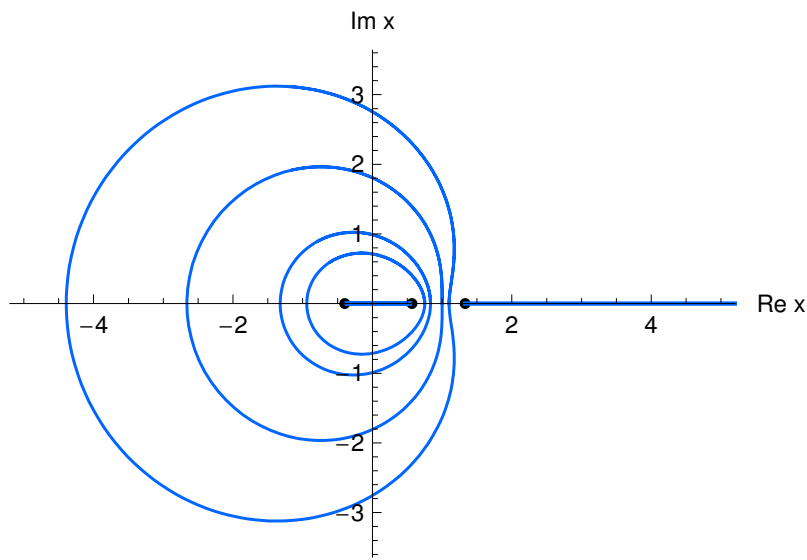}
\caption{Six classical trajectories in the complex-$x$ plane for a particle of
energy $E=0.1$ in the potential $\half x^2-\third x^3$. The three turning points
are located at $x=-0.398,~0.567,~1.331$. There are two classically allowed
regions, one in which a classical particle oscillates between the turning points
at $-0.398$ and $0.567$, and a second that includes the real axis to the right
of $1.331$ in which a classical particle drifts off to infinity in finite time.
All other classical trajectories are closed and periodic.}
\label{f7}
\end{figure*}

Let us examine what happens to a classical particle under the influence of this
potential. First, we choose $g=\frac{1}{3}$ and give this particle a {\it real}
energy $E=0.1$. Because this energy is real, the particle trajectories are
closed and periodic (see Fig.~\ref{f7}). Periodic trajectories cannot represent
a physical tunneling process in which a particle, initially confined in a
potential well, gradually leaks out to infinity.

\begin{figure*}[t!]
\vspace{2.3in}
\includegraphics{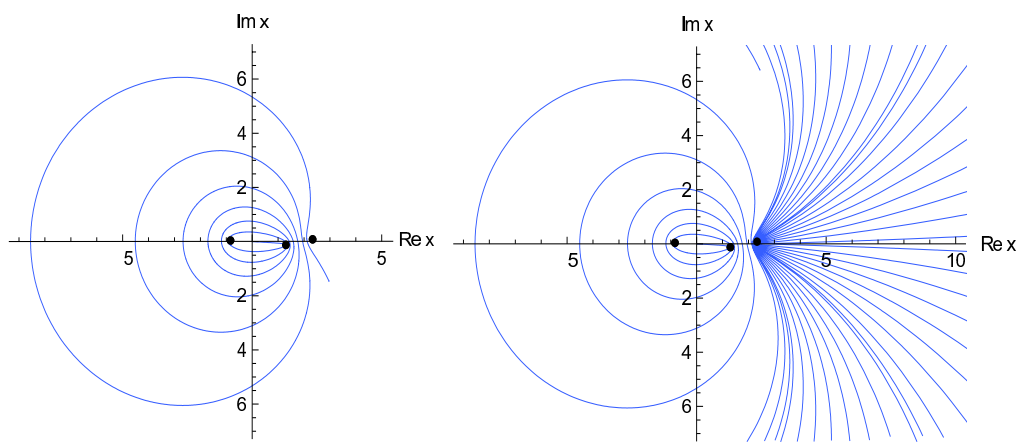}
\caption{Trajectory in the complex-$x$ plane of a classical particle of complex
energy $E=0.456-0.0489i$ in a $\half x^2-2x^3/\sqrt{125}$ potential. The left
trajectory begins at $x=0$ and terminates at $t_{\rm max}=50$, while the right
trajectory runs until $t_{\rm max}=200$. The right turning point takes control
at about $t=40$ and this is in good agreement with the lifetime $\tau$ of the
quantum state, whose numerical value from (\ref{e8}) is about $\tau=20$.}
\label{f8}
\end{figure*}

Next, we allow the energy of the particle to have an imaginary component. We
take $g=2/\sqrt{125}$ and set $E=0.456-0.0489i$. This choice lifts the two outer
turning points slightly above the real-$x$ axis and pushes the central turning
point below the axis, as shown in Fig.~\ref{f8}. The trajectory on the left
represents a classical particle that starts at the origin and goes until $t_{\rm
max}=50$, while the trajectory on the right runs until $t_{\rm max}=200$.
Observe that the classical particle begins its motion by spiraling outward under
the control of the two left turning points, which mark the edges of the
confining region. After some time, the third turning point gradually takes
control. We observe this change of influence as follows: Initially, as the
particle crosses the real axis to the right of the middle turning point, its
trajectory is concave leftward, but as time passes, the trajectory becomes
concave rightward. It is clear that by the fifth orbit the right turning point
has gained control, and we can declare that the classical particle has now
``tunneled'' out and escaped from the parabolic confining potential. The time at
which this classical changeover occurs is approximately at $t=40$. This is in
good agreement with the lifetime of the quantum state in (\ref{e8}), whose
numerical value is about 20.

\section{Periodic potential}
\label{s6}

A periodic potential is used to model the behavior of a quantum particle in a
crystal lattice. When the energy of such a particle lies below the top of the
potential, there are infinitely many disconnected classically allowed regions on
the real axis. Ordinarily, a quantum particle is confined to one such region,
but has a finite probability of tunneling to an adjacent classically allowed
region. Thus, such a particle hops at random from site to adjacent site in the
crystal. However, for a narrow band (or bands) of energy the particle may drift
freely from site to site in one direction. In such a {\it conduction band} the
motion of the particle is said to be {\it delocalized}.

\begin{figure*}[t!]
\vspace{2.9in}
\includegraphics{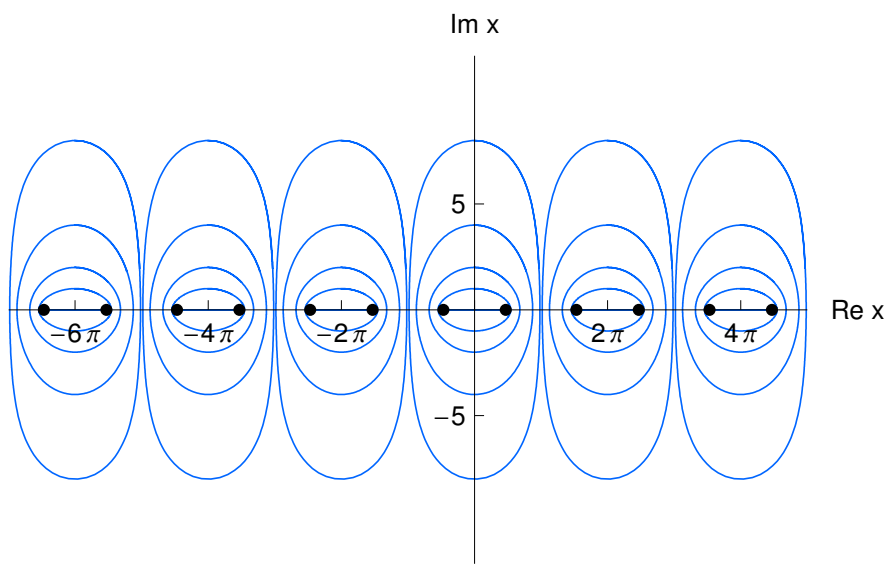}
\caption{Classical trajectories in the complex-$x$ plane for a particle of
energy $E=-0.09754$ in a $-\cos x$ potential. The motion is periodic and the
particle remains confined to a cell of width $2\pi$. Five trajectories are shown
for each cell. The trajectories shown here are the periodic analogs of the
trajectories shown in Fig.~\ref{f4}.}
\label{f9}
\end{figure*}

A classical particle in such a periodic potential exhibits these characteristic
behaviors, but only if its energy is complex. If its energy is taken to be real,
the particle merely exhibits periodic motion and remains confined forever to
just one site in the lattice. Figure \ref{f9} illustrates this periodic motion
for a particle of energy $E=-0.09754$ in a $-\cos x$ potential.

\begin{figure*}[t!]
\vspace{3.3in}
\includegraphics{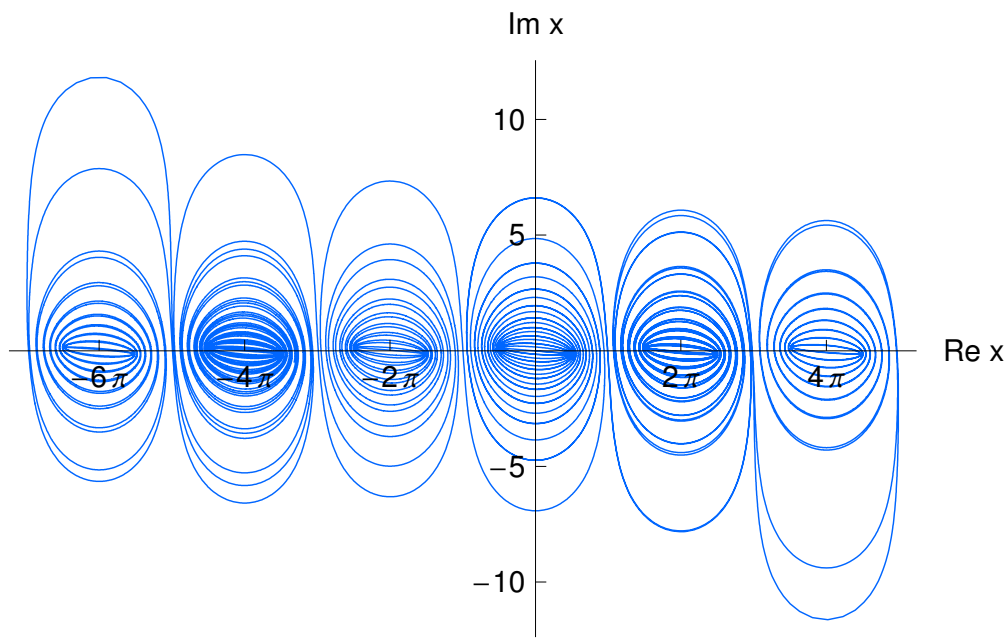}
\caption{Classical trajectory in the complex-$x$ plane of a particle of energy
$E=-0.1-0.15i$ in a $-\cos x$ potential. The particle starts at the origin $x=
0$, spirals outward, and hops to the adjacent site to the right. Then it spirals
inward and back outward and hops to the right again. Then, it hops leftward
five times, and back to the right once more. This deterministic ``random walk''
is reminiscent of a localized quantum particle in a crystal.}
\label{f10}
\end{figure*}

For the same potential, if we take the energy to be complex $E=-0.1-0.15i$, the
classical particle now executes localized hopping from site to site (see
Fig.~\ref{f10}). In this figure the particle starts at the origin $x=0$ and
``tunnels'' right, right, left, left, left, left, left, right, and so on,
without ever crossing its trajectory. This deterministic ``random walk'' is
reminiscent of the behavior of a localized quantum particle hopping randomly
from site to site in a crystal.

\begin{figure*}[t!]
\vspace{2.8in}
\includegraphics{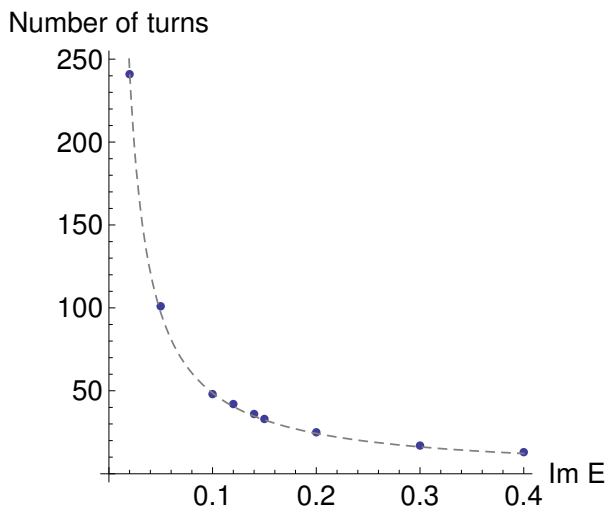}
\caption{Number of turns in the spiral tunneling process shown in Fig.~\ref{f10}
that are required for a classical particle to hop from one site to an adjacent
site versus the imaginary part of its energy. This graph shows that the product
of the tunneling time and the imaginary part of the energy is a constant. For
the time-energy uncertainty principle this product must be greater than $\half$,
and this product is about 17 for the classical $-\cos x$ potential.}
\label{f11}
\end{figure*}

The tunneling rate of the classical particle depends on the imaginary part of
its energy. To measure the time required for the particle in Fig.~\ref{f10} to
hop to an adjacent site, we simply count the number of turns in the spiral path
contained in each cell. This gives an extremely accurate measure of the time
the particle spends in each cell before it hops to an adjacent cell. If we then
vary the imaginary part of the energy and plot the relationship between ${\rm
Im}\,E$ and the tunneling time, we obtain the graph shown in Fig.~\ref{f11}. We
see from this graph that the product of ${\rm Im}\,E$ and the tunneling time is
a constant. In units where $\hbar=1$, for the time-energy quantum uncertainty
principle this product should be greater than $\half$, and this lower bound is
saturated by the harmonic oscillator. Measured for the $-\cos x$ potential, we
find that the numerical value of this product is approximately $17$.

\begin{figure*}[t!]
\vspace{2.2in}
\includegraphics{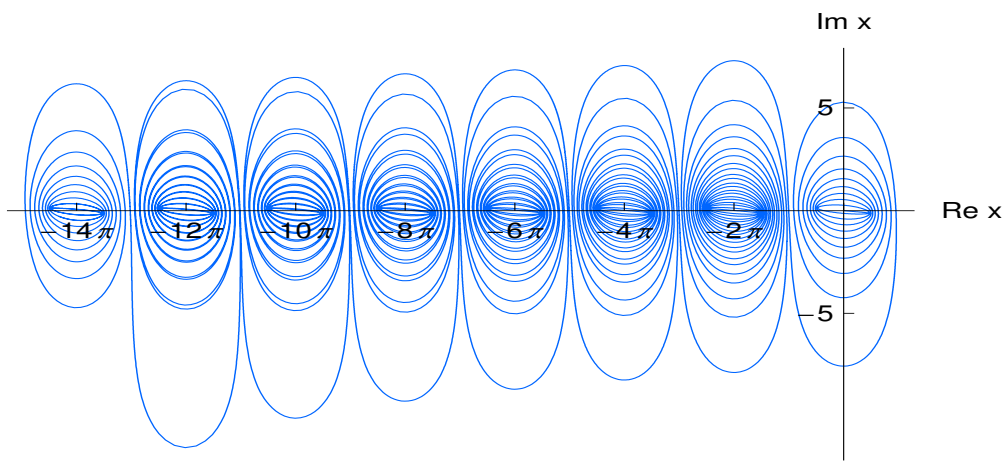}
\caption{Classical trajectory in the complex-$x$ plane for a particle of energy
$E=-0.09754-0.1278i$ in a $-\cos x$ potential. The particle starts at $x=0$ and
behaves like a delocalized quantum particle in a conduction band. It drifts
leftward at a nearly constant rate, spiraling inward about twenty times and then
outward about twenty more times before crossing to the next adjacent cell.}
\label{f12}
\end{figure*}

We have found that there is a narrow range for the real part of the energy in 
which the classical particle in the $-\cos x$ potential behaves as if it is a
delocalized quantum particle in a conduction band; that is, the classical
particle drifts consistently from site to site in the potential in one
direction. A classical trajectory that illustrates this behavior is shown in
Fig.~\ref{f12}. The energy of this particle is $E=-0.09754-0.1278i$. The band
edges can be determined with great numerical precision; we find that when ${\rm
Im}\,E=-0.1278$, the range of real energy in this band is $-0.1008<{\rm Re}\,E<
-0.0971$.

\section{Conclusions}
\label{s7}

The classical differential equations that we have solved numerically in this
paper can be solved exactly in terms of elliptic functions, but we have not done
so because we are interested in their qualitative behavior only. Their solutions
seem to suggest that many features thought to be only in the quantum arena can
be reproduced in the context of complex classical mechanics. Of course, we have
not shown that {\it all} quantum behavior can be recovered classically. In
particular, we have not yet been able to observe the phenomenon of interference.
(For example, we do not yet see the analog of the nodes in bound-state
eigenfunctions.)

The ideas discussed here might be viewed as a vague alternative version of a
hidden-variable formulation of quantum mechanics. The original idea of
de~Broglie, Bohm, and Vigier was that a quantum system can be reduced to a
deterministic system in which probabilities arise from the lack of knowledge
concerning certain hidden variables. This approach encountered various
difficulties, but an alternative way forward was suggested by Wiener and Della
Riccia \cite{R13,R14}, who argued that the hidden quantity in a coordinate-space
representation of quantum mechanics is not the classical position of the
particle but rather the momentum variable, which is integrated out and thus
circumvents the issues associated with traditional approaches. In order to
obtain the quantization condition, Wiener and Della Riccia introduced
probability distributions over the classical phase space, thus obtaining the
spectral resolution of the Liouville operator in terms of the eigenfunctions of
the associated Schr\"odinger equation. The complex energy formulation of
classical mechanics outlined here, if viewed as an alternative hidden variable
theory, is close in spirit to that of Wiener and Della Riccia, but is distinct
and more primitive in that we have not introduced probability distributions over
the classical phase space. Nevertheless, the inaccessibility of the imaginary
component of the energy in classical mechanics might necessitate introducing a
probability distribution for the energy, which in turn might give rise to a more
precise statement of uncertainty principles. The analogies between quantum
mechanics and complex-energy classical mechanics reported here make further
investigations worthwhile.

\vspace{0.5cm}
\footnotesize
\noindent
We thank M.~Berry, G.~Dunne, H.~F.~Jones, and R.~Rivers for helpful discussions.
CMB is supported by a grant from the U.S. Department of Energy.

\end{document}